\documentclass[pra,twocolumn,showpacs,amsmath,amssymb]{revtex4}

\usepackage[dvips]{color}
\usepackage{graphicx}

\begin{document}

\title{Finite temperature QMC study of the one-dimensional polarized
  Fermi gas}

\author{M. J. Wolak$^1$, V. G. Rousseau$^2$,
  C.~Miniatura$^{3,1,4}$, B.~Gr\'emaud$^{5,1,4}$, R. T. Scalettar$^6$,
  G. G. Batrouni$^{3,1}$}

\affiliation{
$^1$ Centre for Quantum Technologies,
National University of Singapore; 2 Science Drive 3 Singapore 117542\\ 
$^2$Department of Physics and Astronomy, Louisiana State
  University, Baton Rouge, Louisiana 70803, USA\\
$^3$INLN, Universit\'e de Nice--Sophia
Antipolis, CNRS; 1361 route des Lucioles, 06560 Valbonne, France\\ 
$^4$Department of Physics, National University of Singapore, 2 Science
Drive 3, Singapore 117542, Singapore\\
$^5$ Laboratoire Kastler Brossel, UPMC-Paris 6, ENS, CNRS; 4 Place
Jussieu,F-75005 Paris, France\\
$^6$Physics Department, University of California, Davis, California
95616}

\begin{abstract}
Quantum Monte Carlo (QMC) techniques are used to provide an
approximation-free investigation of the phases of the one-dimensional
attractive Hubbard Hamiltonian in the presence of population
imbalance.  The temperature at which the
``Fulde-Ferrell-Larkin-Ovchinnikov'' (FFLO) phase is destroyed by
thermal fluctuations is determined as a function of the polarization.
It is shown that the presence of a confining potential does not
dramatically alter the FFLO regime, and that recent experiments on
trapped atomic gases likely lie just within the stable temperature
range.
\end{abstract}


\maketitle


\section{Introduction}

Pair formation between fermions is one of the most rich and active
fields of investigation in condensed matter systems, occuring in
contexts as varied as exciton formation in quantum well structures
\cite{golub} to its most well-studied realization, the superconducting
state \cite{BCS}.  Indeed, the possibility of pair formation has many
even wider implications, including the early speculation that it might
dominate the nature of interior of neutron stars \cite{ginzburg}.  By
far the most commonly studied situation is one in which the
populations of the two fermion species (normally up and down spins)
are balanced, and Cooper pairs form with zero center of mass momentum
\cite{cooper}. Soon after the development of the BCS theory \cite{BCS}
of superconductivity, the question of pair formation in polarized
superconducting systems, {\it i.e.} when the populations of the two
spin states are imbalanced, was addressed independently by Fulde and
Ferrel \cite{fulde} (FF), Larkin and Ovchinnikov \cite{larkin} (LO)
and Sarma \cite{sarma}. The question was motivated by interest in the
nature of superconductivity in the presence of a magnetic field, which
induces spin polarization.

FF and LO (henceforth FFLO) proposed similar, yet not identical,
mechanisms whereby the Cooper pairs form with non-zero center of mass
momentum equal to the difference of the now-unequal Fermi momenta of
the two species. As a consequence, in the FFLO scenario the order
parameter is not homogeneous; the system has pair-rich regions
separated by regions depleted in pairs but with an excess of the
majority, unpaired fermion species. In the competing proposal by
Sarma, the Fermi surfaces deform to allow pair formation with zero
center of mass momentum. This means that the system remains uniform;
it is a homogeneous mixture of pairs and unpaired fermions from the
majority population. In other words, in the FFLO mechanism, the
momentum distribution of pairs has its peak at a momentum equal to the
difference between the two Fermi momenta, $k_{\rm
  peak}=|k_{F1}-k_{F2}|$. On the other hand, the Sarma mechanism would
result in a pair momentum distribution with $k_{\rm peak}=0$.
Deciding between these two scenarios experimentally proved to be very
difficult; the observation of the FFLO phase in solids was only
achieved relatively recently in heavy fermion systems
\cite{radovan03}.

This question has recently taken on added importance. Interest in the
astrophysical community has continued to grow and broaden.  It is now
believed that at extreme conditions of pressure, for example in the
interior of supermassive stars, quark matter forms and pairing between
the quarks could lead to color superconductivity
\cite{Casalbuoni}. More immediate experimental systems where these
effects may be observed, and are indeed sought, are ultra-cold atoms
confined in traps where two hyperfine states of fermionic atoms play
the role of up and down spins.  Such experiments have now reported the
presence of pairing in the case of unequal populations
\cite{zwierlein06,partridge06} in three-dimensional cigar shaped traps
and in one dimensional traps \cite{hulet}. However, the precise nature
of the pairing has not yet been elucidated experimentally.

On the theoretical side, much effort has been put into understanding
the pairing mechanism. Calculations using mean field theory
\cite{castorina05,sheehy06,kinnunen06,machida06,gubbels06,parish07,hu07,he07,wilczek,koponen,Drummond1},
effective Lagrangian \cite{son06} and Bethe ansatz \cite{orso} studies
have been performed for the uniform system, with extensions to the
trapped system using the local density approximation (LDA).  Extensive
numerical work for the one dimensional system using Quantum Monte
Carlo (QMC) \cite{batrouni08,casula}, and the Density Matrix
Renormalization Group (DMRG) \cite{feiguin07,luscher,rizzi,tezuka07})
has demonstrated that, in the ground state, population imbalance leads
to a robust FFLO phase over a very wide range of polarization and
interaction strengths.

In addition to the above case of population imbalance, the case of
mass imbalance has been addressed theoretically
\cite{wilczek1,wilczek2} and numerically \cite{batrouni09}.  This case
is relevant to color superconductivity, where the quark masses are not
equal, and also to ultra-cold $^{40}$K-$^6$Li atomic mixtures.

The stability of the FFLO phase at finite temperatures remains an open
question. Mean field calculations
\cite{KoponenfiniteT,DrummondfiniteT,Kakashvili} possibly shed some
light but can be less reliable in low dimension where quantum
fluctuations are large. This question is of paramount importance for
experiments, especially in trapped atomic systems, since difficulties
in cooling fermionic atoms raise concerns about whether the currently
attainable temperatures are low enough for a thorough investigation of
FFLO physics.  The present paper reports on QMC studies of this issue
which provide an exact treatment of interaction effects on lattices of
finite size.  The key result is that, when the polarization is
sufficiently large, the paired phase can exist up to temperatures of
order one tenth of the Fermi energy, even in the presence of a
confining potential.  Current experiments on trapped atoms are likely
in this temperature range.

The paper is organized as follows. In the next section we will discuss
the model and the numerical methods used. In section III, we discuss
our results for the phase diagram of the uniform one-dimensional
system at finite temperature. In section IV we discuss the trapped one
dimensional system followed by our conclusions in section V.

\section{Model and Methods}
In order to study the pairing mechanism of fermions in an optical
lattice, we consider the one-dimensional fermionic Hubbard
Hamiltonian,
\begin{eqnarray}
\label{Hamiltonian}
H&=&-t\sum_{i\, \sigma} (c_{i\,\sigma}^{\dagger}
c_{i+1\,\sigma}^{\phantom\dagger} + c_{i+1\,\sigma}^\dagger c_{i \,
\sigma}^{\phantom\dagger}) - \sum_i ( \mu_1\hat n_{i\,1} +
\mu_2 \hat n_{i\,2}) \nonumber \\ 
\nonumber
&&+U \sum_{i} \left(\hat n_{i\,1}-\frac{1}{2}\right)\left(\hat
n_{i\,2}-\frac{1}{2}\right)\\ 
&&+V_T \sum_{i} \left(x_i-\frac{L}{2}\right)^2 \left(\hat n_{i \,1} +
\hat n_{i \, 2}\right) 
\end{eqnarray}
where $c_{i\,\sigma}^\dagger$ and $c_{i\, \sigma}^{\phantom\dagger}$
are fermion creation and annihilation operators on lattice site $i$
satisfying the usual anticommutation relation,
$\{c^{\phantom\dagger}_{i\,\sigma},c^\dagger_{j\,
  \sigma^{\prime}}\}=\delta_{i,j}\delta_{\sigma,\sigma^\prime}$. The
fermionic species are labeled by $\sigma=1,2$ and $\hat
n_{i\,\sigma}=c_{i\,\sigma}^\dagger c_{i\, \sigma}^{\phantom\dagger}$
is the corresponding number operator. The energy scale is set by
taking the hopping parameter $t=1$.  The contact interaction strength
is $U$ which is negative since we are interested in pair formation in
the attractive model. In the grand canonical ensemble, the populations
are fixed by tuning the chemical potentials for the two species,
$\mu_1$ and $\mu_2$, whereas in the canonical ensemble this term is
absent and the populations are fixed simply by choosing the number of
particles. As we discuss below, both ensembles will be used in this
work. The last term describes the confining harmonic trap which is
centered at the midpoint, $L/2$, of the $L$-site lattice. We take
periodic boundary conditions.

Our Quantum Monte Carlo (QMC) results are obtained using two different
methods: the Determinant QMC algorithm \cite{DQMC} (DQMC) and the
Stochastic Green Function (SGF) technique \cite{SGF,DirectedSGF}. In
DQMC, the interaction term in the Hamiltonian, which is quartic in the
fermionic operators, is decoupled via the Hubbard-Stratonovich (HS)
transformation. This allows the fermion operators to be traced out,
giving an expression for the partition function as an integral over
the configurations of the auxiliary HS field.  We employ the grand
canonical formulation of DQMC.  The statistical weight of these
configurations is given by the product of the spin up and down
determinants, each of which can change sign depending on the HS
configuration. When the populations are balanced, $\mu_1=\mu_2$, it is
well known that DQMC does not suffer from the sign problem for an
attractive interaction ($U<0$): The two determinants are identical and
thus their product is always positive. This is no longer the case when
the populations are imbalanced, $\mu_1 \neq \mu_2$ (or for $U>0$),
resulting in the appearance of the sign problem. The severity of this
problem depends on several factors such as the size of the system, the
coupling constant, the temperature and the imbalance in the
populations. For the systems and parameters we consider in this work
(see below) the average sign did not fall below $0.5$, which allowed
us to obtain good statistical precision in our simulations. We used
this algorithm for the uniform system; it has the advantage of rather
fast convergence rates, but to perform simulations at fixed particle
numbers, the chemical potentials need to be tuned. A typical
simulation for $L=50$ sites and $\beta=30$ with an imaginary time step
$\Delta \tau=0.1$ runs for about $4$ to $5$ days on a desktop computer
and yields error bars of the order of $0.5\%$ for the peak of the pair
momentum distribution, a central quantity of interest in this work.

The SGF algorithm \cite{SGF} with directed update \cite{DirectedSGF}
can be used both in the grand canonical or canonical modes.  We
describe it in more detail than DQMC since it has been developed much
more recently.  First, the Hamiltonian is written as $ H=\hat{\cal
  V}-\hat{\cal T}$, where $\hat{\cal V}$ is diagonal in the occupation
number basis of the direct space, and $\hat{\cal T}$ is the remaining
non-diagonal part. The algorithm samples an extended partition
function represented in the interaction picture,
\begin{equation}
  \label{ExtendedZ} 
{\cal Z}(\beta,\tau)={\rm Tr}\,e^{-\beta\hat{\cal
    V}}T_\tau\Big[\hat{\cal G}(\tau)e^{\int_0^\beta\hat{\cal
    T}(\tau)d\tau}\Big], 
\end{equation}
where $T_\tau$ is the time-ordering operator and the time dependence
of any operator $\hat{\cal A}(\tau)$ is given by the interaction
representation $\hat{\cal A}(\tau)=e^{\tau\hat{\cal V}}\hat{\cal A}
e^{-\tau\hat{\cal V}}$. The ``Green operator" $\hat{\cal G}$ is
defined by its matrix elements, $\big\langle\psi_L\big|\hat{\cal
  G}\big|\psi_R\big\rangle=g_{pq}$, where $p$ and $q$ are the number
of creations and destructions of particles that transform the
(occupation number) state $\psi_R$ into $\psi_L$. The matrix $g_{pq}$
is a decreasing function of $(p+q)$ \cite{SGF}.  The extended
partition function (\ref{ExtendedZ}) is expanded in $\hat{\cal T}$
yielding operator strings of the form:
\begin{equation}
  \label{OperatorString} 
\cdots\hat{\cal T}(\tau_{L+2})\hat{\cal T}(\tau_{L+1})\hat{\cal
  T}(\tau_{L})\hat{\cal G}(\tau)\hat{\cal T}(\tau_{R})\hat{\cal
  T}(\tau_{R-1})\hat{\cal T}(\tau_{R-2})\cdots, 
\end{equation}
The Green operator updates the operator string (\ref{OperatorString})
by propagating in imaginary time, while inserting and removing
$\hat{\cal T}$ operators. From its definition, the Green operator
contains terms that do not conserve the number of particles. However
if the Hamiltonian commutes with the operator that measures the total
number of particles, $\hat{\cal N}$, for example the Hamiltonian
(\ref{Hamiltonian}), then only conservative terms of $\hat{\cal G}$
have non vanishing contributions since the trace imposes the same
number of particles both at the begining and the end of the operator
string. As a consequence the number of particles remains strictly
constant and the SGF algorithm works in the canonical ensemble by
nature.  However, a simple trick can be used in order to simulate
exactly the grand-canonical ensemble. The idea is to add a non
conservative part $\hat{\cal H}_{\textrm{nc}}$ to the Hamiltonian,
\begin{equation}
\label{NonConservative} 
\hat{\cal H}_{\textrm{nc}}=\gamma\sum_j
  \big(a_j^\dagger+a_j^{\phantom\dagger}\big), 
\end{equation}
where $\gamma$ is an optimization parameter, and allows the Green
operator to insert at most one $\hat{\cal H}_{\textrm{nc}}$ operator
in the string. This allows the number of particles to fluctuate, while
the addition of the usual term $-\mu\hat{\cal N}$ to the Hamiltonian
determines the mean number of particles via the chemical potential
$\mu$. When measuring physical quantities, ignoring configurations in
which the operator string contains a $\hat{\cal H}_{\textrm{nc}}$
operator, corresponds to integrating over these configurations the
probability of going from a given configuration with $N$ particles and
no $\hat{\cal H}_{\textrm{nc}}$ to another one with $M$ particles and
no $\hat{\cal H}_{\textrm{nc}}$, via intermediate configurations with
$\hat{\cal H}_{\textrm{nc}}$. This integrated probability corresponds
exactly to the probability of going from one configuration with $N$
particles to another one with $M$ particles. As a result, the
configurations of the grand-canonical partition function $\textrm{Tr
}\,e^{-\beta(\hat{\cal H}-\mu{\hat N})}$ are generated with the
correct Boltzmann weight.

To use this algorithm to simulate fermions in one dimension, we first
use the Jordan-Wigner transformation to map the system onto a system
of hard core bosons. Consequently, this algorithm does not suffer from
the sign problem but, clearly, it cannot be used in higher dimensions
where the Jordan-Wigner transformation fails to solve the sign
problem. We used this algorithm in the canonical ensemble mainly for
the confined system where it is very convenient to control the
populations directly rather than tune the chemical potentials. A
typical simulation for $L=120$ and $\beta=32$ runs for about $5$ days
on a desktop computer and yields an error of the order of $0.5\%$ for
the peak of the pair momentum distribution..

As part of our code verification, the grand canonical SGF and DQMC
were compared for the same parameters and found to yield the same
results within the error bars. Furthermore, we verified that DQMC and
grand canonical SGF both agreed with the canonical SGF when the
chemical potentials in the grand canonical cases were tuned to give
the same populations as the canonical cases, see for example
Ref.[\onlinecite{batrouni08}]. The results presented below were
obtained with these three algorithms; the choice being dictated by the
specific measurement we were after. Because of the equivalence of the
three algorithms, we will not comment below on which results were
obtained with which algorithm.

Using these algorithms, we calculate both the real space Green
functions of the species, $G_\sigma$, and the pair green function,
$G_{\rm pair}$,  
\begin{eqnarray}
\label{gfct}
G_\sigma(l) &=& \langle c_{j+l\,\sigma}^\dagger c_{j
\,\sigma}^{\phantom\dagger} \rangle,\\
\label{pairgfct}
G_{\rm pair}(l) &=& \langle
\Delta^{\dagger}_{j+l}\,\Delta_{j}^{\phantom\dagger} \rangle,\\ 
\label{pairoperator}
\Delta_j &=& c_{j \, 2} \,c_{j \, 1},
\end{eqnarray}
where $\Delta_j$ destroys a pair on site $j$. The Fourier transform of
$G_{\sigma}(l)$ yields the momentum distributions $n_{\sigma}(k)$ and
the transform of $G_{\rm pair}(l)$ leads to the pair momentum
distribution, $n_{\rm pair}(k)$, a central quantity in this work. In
the non-interacting limit, the Fermi momentum of a population is given
by $k_{F\sigma}=\frac{N_{\sigma}-1}{2}\frac{2\pi}{L}$, where $L$ is
the length of the system and $N_{\sigma}$ the number of particles.

\section{Phase diagram of the homogenous system}

  We begin with a discussion of the physics in the absence
of a confining potential, $V_T = 0.$

As discussed in the introduction, it is now generally agreed that the
ground state of a Fermi system with attractive interactions and
imbalanced populations is the FFLO state. The mismatch in the Fermi
momenta results in pair formation with nonzero center-of-mass momentum
$k=\pm |k_{F_1}-k_{F_2}|$. Consequently the pair momentum
distribution, the Fourier transform of the pair Green function
Eq.(\ref{pairgfct}), peaks at this value of the momentum. This peak at
nonzero momentum serves as the principal diagnostic indicating the
presence of the FFLO
state \cite{batrouni08,luscher,rizzi,tezuka07,feiguin07}.

The situation at finite temperature, which is important
experimentally, is less clear. Approximate methods, such as mean
field, do not always yield the same phase diagram. In this section we
will map out the phase diagram in the polarization-temperature plane
where the polarization is defined by,
\begin{equation}
P=\frac{N_1-N_2}{N_1+N_2},
\label{polarization}
\end{equation}
where $N_1$ ($N_2$) is the majority (minority) population and
$N=N_1+N_2$ is the total number of particles. To this end, we study,
at fixed $P$, the behaviour of the pair momentum distribution, $n_{\rm
pair}(k)$, as a function of the temperature $T$. For very low $T$,
$n_{\rm pair}(k)$ peaks at $k\neq 0$ and the system is in the FFLO
state. As $T$ is increased, the peak in $n_{\rm pair}$ gets lower and
shifts to $k=0$. The temperature at which this first happens is the
cross-over temperature, $T_c$. Note that in this one-dimensional
system, transitions at finite temperature are not true phase
transitions, but rather cross-overs.

\begin{figure}[h]
\begin{center}
 \includegraphics[width=0.45\textwidth,angle=0,clip]{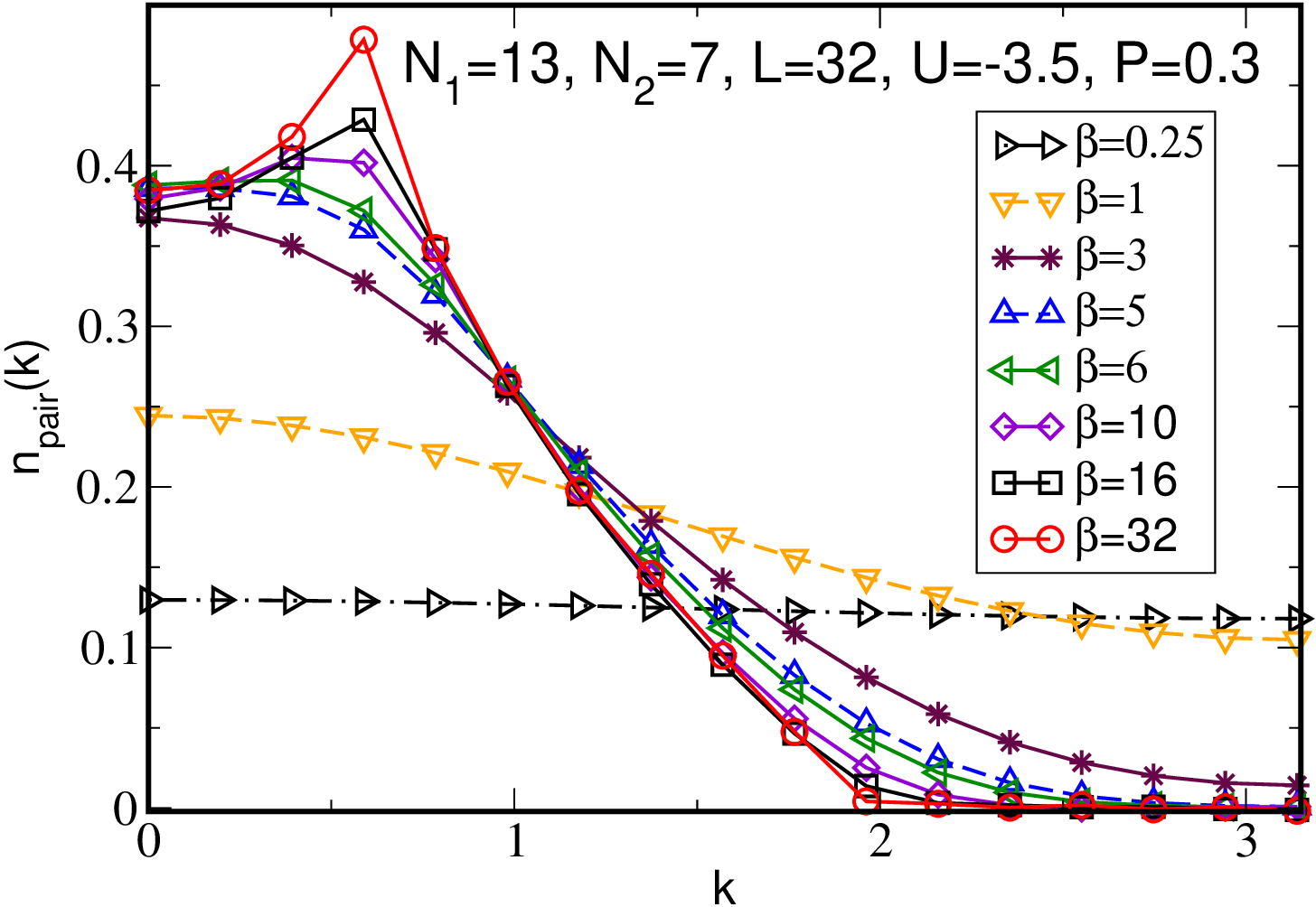}
\end{center}
\caption{\label{fig:finiteTpairmom} (color online) The effect of
  temperature on the pair momentum distribution. A peak at non-zero
  momentum is a signature of the FFLO state. As $\beta$ decreases, the
  peak disappears at a crossover value $\beta_c=1/T_c$. In this case,
  $\beta_c\approx 6$. The error bars are of the order of the symbol
  size.}
\end{figure}

This is illustrated in Fig.~\ref{fig:finiteTpairmom} where we show QMC
results for $n_{\rm pair}(k)$ as a function of $k$ for several values
of the inverse temperature $\beta$. The simulations were done for
fixed populations, $N_1=13$ and $N_2=7$ on a system with $L=32$
lattice sites and an attractive interaction $U=-3.5t$. The figure
shows clearly that as $\beta$ decreases from $\beta=32$, the height of
the FFLO peak decreases and, in fact, shifts to lower $k$ values. The
shift to lower $k$ values is made more evident by simulating larger
systems since this gives more $k$ grid points. When the peak at
nonzero $k$ is equal to $n_{\rm pair}(0)$ to within $1\%$, we consider
the peak to have shifted to $k=0$ and the FFLO state to have
disappeared. In the Fig.~\ref{fig:finiteTpairmom} this happens for
$\beta_c\approx 6$. Reducing $\beta$ further leads to continued
decrease of the height of the peak, which remains at $k=0$.

The question then arises as to how the FFLO peak behaves as a function
of $U$ at fixed $\beta$, $N_1$ and $N_2$ (and consequently fixed $P$).
Clearly, for $U=0$ there is no pairing and no FFLO peak; then, as $|U|$
is increased, the peak at nonzero $k$ forms and its height increases.
However, we found that as $|U|$ continues to increase, the FFLO peak
will reach a maximum height and then start to decrease. We also found
that the peak is more sensitive to $|U|$ at higher $T$. We believe
the reason the peak starts to go down at high values of $|U|$ is that
with increasing attraction, the pairing becomes increasingly localized
in space and eventually the paired fermions form a very tightly bound
bosonic molecule and the system resembles closely a usual Bose-Fermi
mixture which does not exhibit FFLO peaks.

\begin{figure}[h]
\begin{center}
 \includegraphics[width=0.45\textwidth,angle=0,clip]{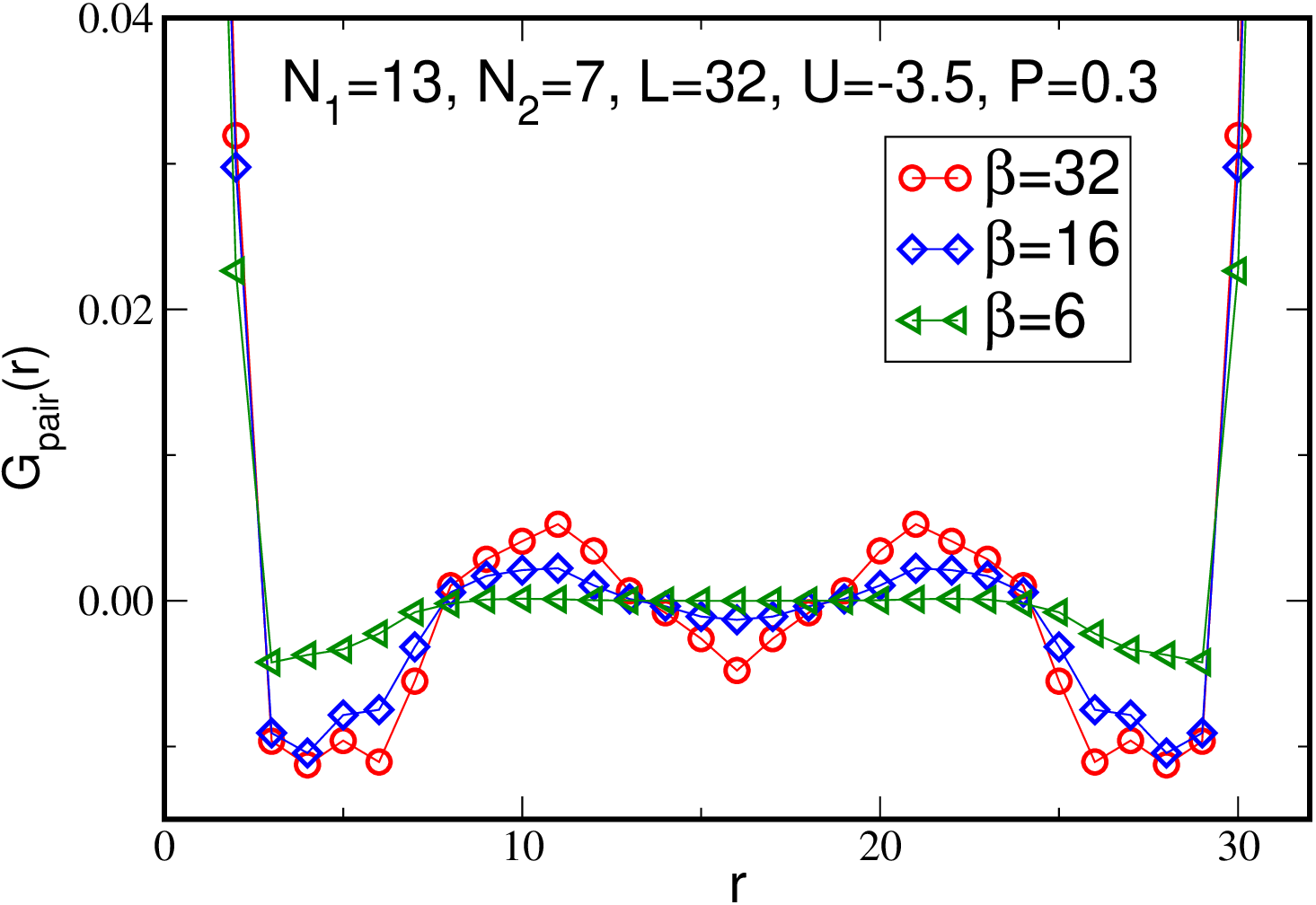}
\end{center}
\caption{\label{fig:finiteTgreenF} (color online) Pair Green function
  calculated at low temperatures where the system is in the FFLO state
  ($\beta=32$ and $\beta=16$) and at the temperature at which FFLO
  disappears ($\beta=6$). Note that the oscillations disappear at the higher
temperature.}
\end{figure}

The peak of $n_{\rm pair}(k)$ at $k_{\rm peak}\neq0$ means that the
system is, in fact, not homogenous: The spatial pair Green function,
Eq.(\ref{pairgfct}), oscillates as a function of distance with a
period given by $2\pi/|k_{\rm peak}|$. These oscillations have been
discussed, for example, in the context of mean field
theory \cite{Drummond1, Trivedi}. Physically, they indicate that the
system has regions which are rich in pairs separated by regions poor
in pairs but rich in the excess fermion species.  Such oscillations
are shown in Fig.~\ref{fig:finiteTgreenF} for three values of
$\beta$. It is seen that as the temperature increases and the height
of the FFLO peak decreases, the oscillations decrease in amplitude and
eventually disappear as homogeneity is restored in the system.

\begin{figure}[h]
\begin{center}
 \includegraphics[width=0.45\textwidth,angle=0,clip]{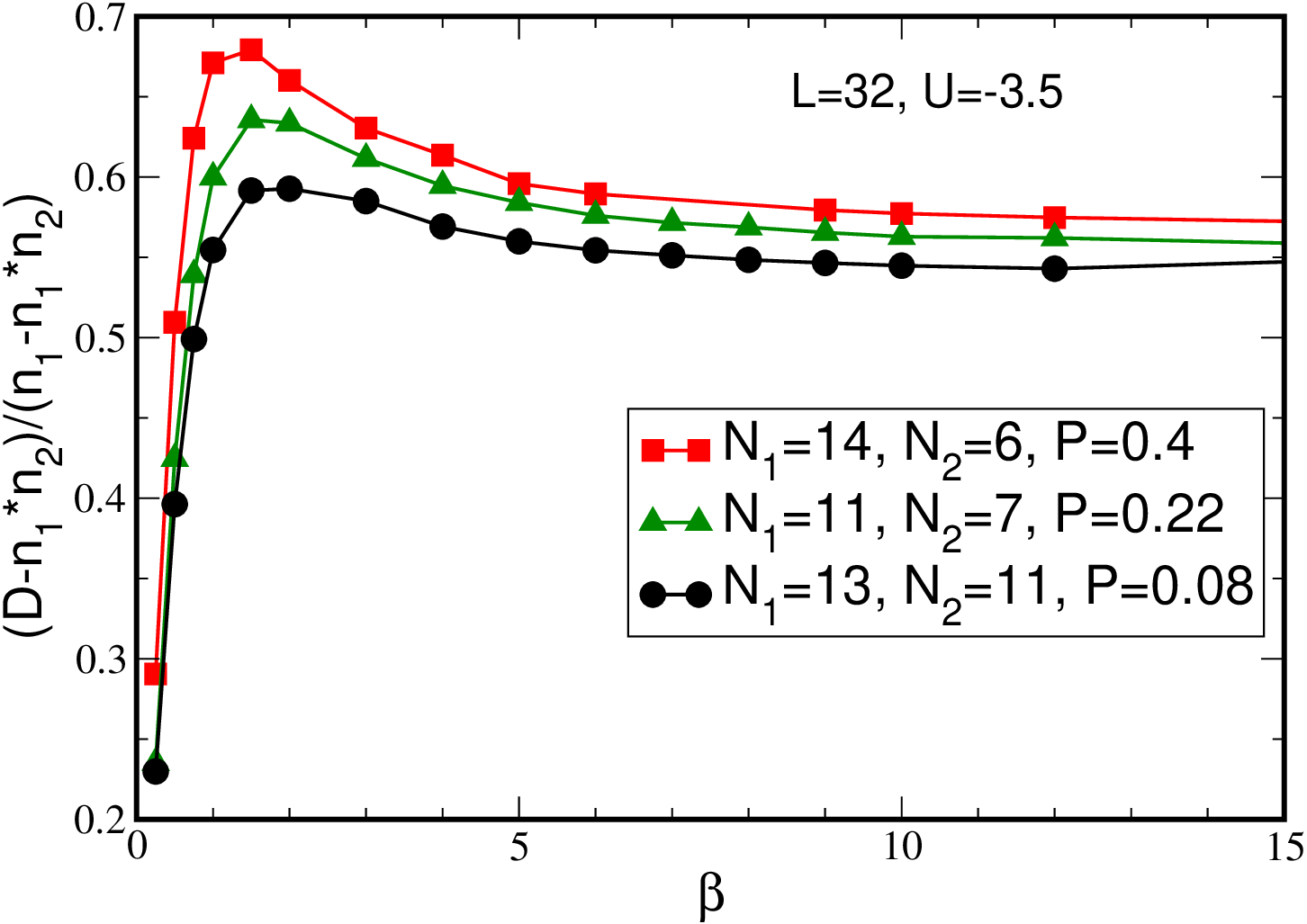}
\end{center}
\caption{\label{fig:pairing} (color online) Double occupancy versus
  $\beta$ for three polarizations exhibits a very sharp drop for
  $\beta<1$ indicating that pairs are being broken near $\beta\sim
  1/|U|$.}
\end{figure}

The question then arises as to the nature of the phase at $T>T_c$. Two
possibilities are: (1) At $T>T_c$ the pairs are broken and the system
is a mixture of two Fermi liquids or (2) pairs are still present but
the system has been homogenized by thermal agitation. A first
indication is given by the energy scales involved. The binding energy
of the pairs at very low temperature is of the order of $|U|$ and in
our system here $|U|=3.5t$. So, to break the pairs, an equivalent
amount of thermal energy is needed which means $\beta\approx
1/|U|$. For the case discussed in Fig.~\ref{fig:finiteTgreenF}, the
crossover from FFLO to the uniform phase happens at $\beta_c\approx 6$
not $t/|U|=0.286$. This means that $T_c$ is more than an order of
magnitude smaller than the temperature needed to break the
pairs. This, then, favors the conclusion that when FFLO first
disappears, the pairs have not yet been broken and the system is in a
homogeneous polarized paired phase (PPP). Another piece of evidence is
provided by studying the average double occupancy of the sites given
by
\begin{equation}
\label{doubleocc}
D =\langle n_{i1} n_{i2}\rangle =
\langle\Delta_{i}^{\dagger}\,\Delta^{\phantom\dagger}_{i}\rangle.
\end{equation}
In the absence of pairing, $\langle n_{i1}n_{i2}\rangle = \langle
n_{i1}\rangle\langle n_{i2}\rangle=N_1N_2/L^2$ while if pairing is
perfect, {\it i.e.} if all the minority particles are paired, $\langle
n_{i1}n_{i2}\rangle =N_2/L$.  We define the normalized double
occupancy by
\begin{equation}
\label{normdoubleocc}
{\cal D}=\frac{D-n_1n_2}{n_2-n_1n_2},
\end{equation}
where $n_1=N_1/L$ and $n_2=N_2/L$ and we recall that $N_2<N_1$. With
this normalization, we have $0\leq {\cal D} \leq 1$. This quantity is
shown in Fig.~\ref{fig:pairing} for three polarizations.  One can see
that the pairing drops significantly at rather high temperatures,
$\beta\approx1/U$. Thus we conclude that the pairs are not broken when
the FFLO peak disappears but the systems is in a PPP. As one can see,
this pairing parameter does not saturate for the case we
presented. One could expect it to reach the maximum value at a very
strong U limit in the low temperature regime, where both thermal
and quantum fluctuations are absent.

\begin{figure}[h]
\begin{center}
  \includegraphics[width=0.45\textwidth,angle=0,clip]{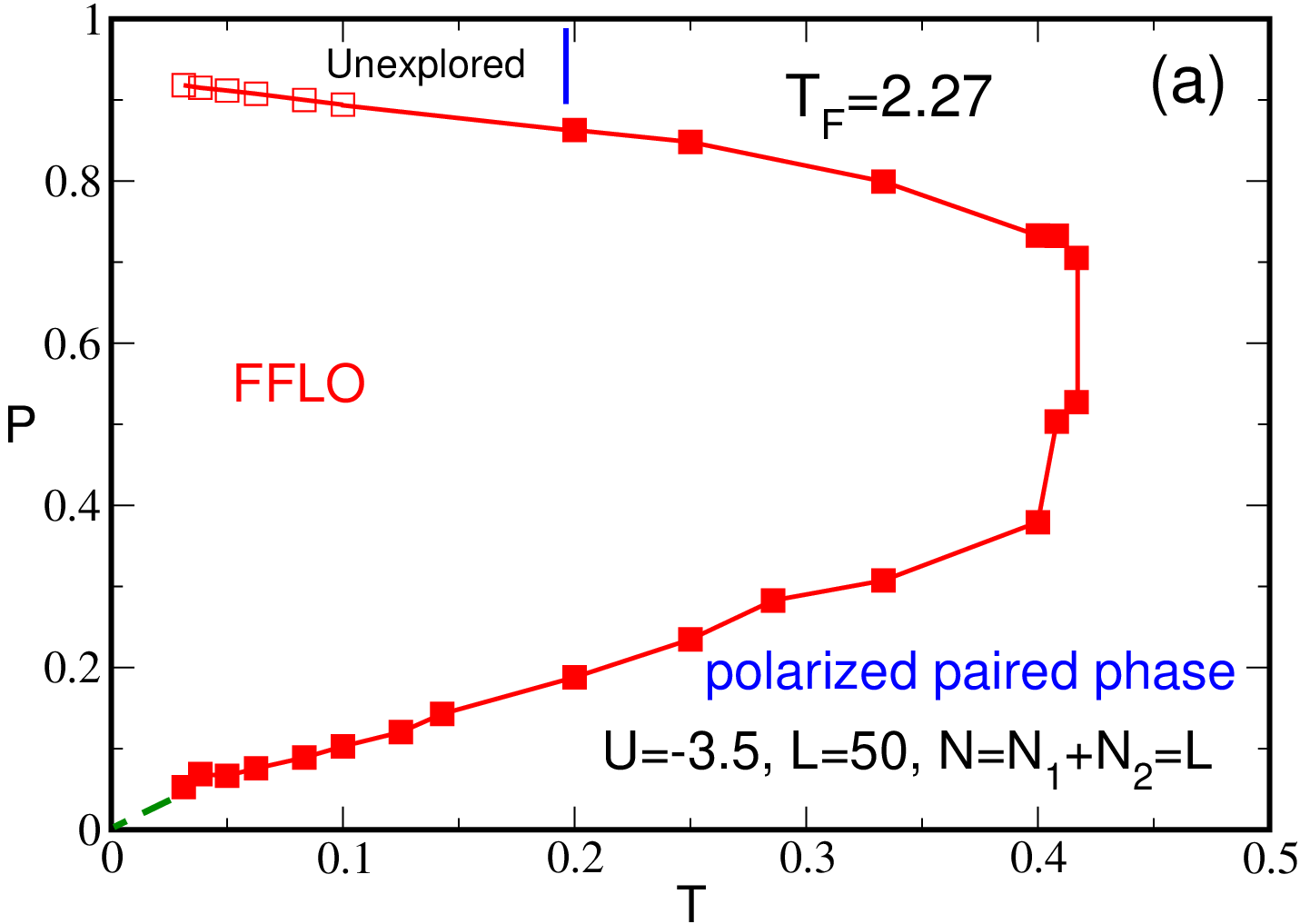}
\end{center}
\begin{center}
  \includegraphics[width=0.45\textwidth,angle=0,clip]{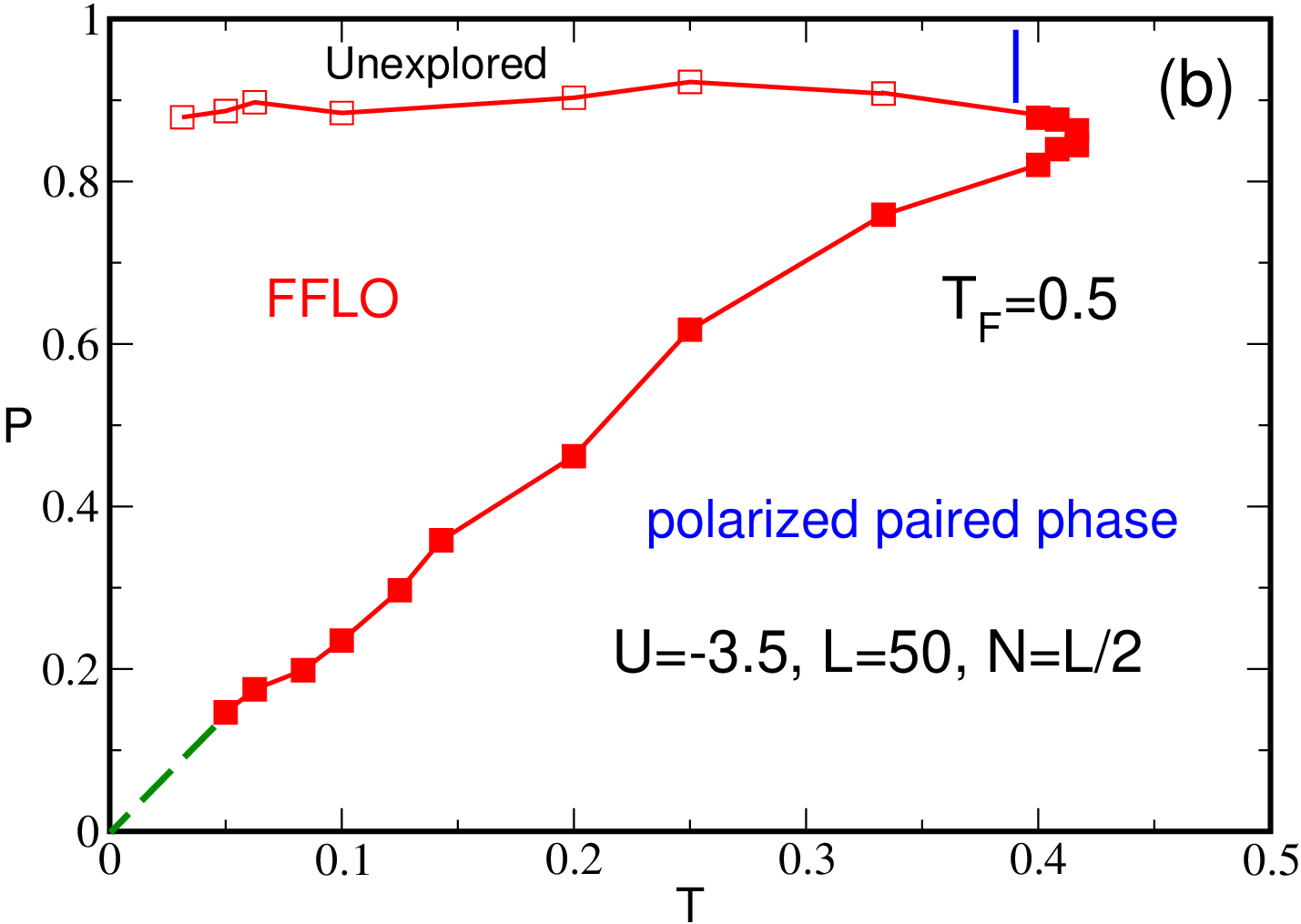}
\end{center}
\vspace{-0.6cm}
\caption{\label{fig:PhaseDiag} (color online). Phase diagram in the
  polarization-temperature plane. (a) the system at $N=L$ (half
  filling), (b) the system at $N=L/2$ (quarter filling). At P=0 system
  is in the BCS state.  The regions above the open (red) squares and
  to the left of the vertical (blue) lines are unexplored.  Up to the
  open squares the system is in the FFLO phase. Phase boundaries
  represent cross-over behaviour not phase transitions. In the FFLO
  phase, $n_{\rm pair}(k)$ peaks at $k\neq 0$, in the PPP phase, the
  peak is at $k=0$.}
\end{figure}

We are now ready to apply the above considerations to determine the
phase diagram in the $(P,T)$ plane. To this end, we keep the total
population, $N$, constant and for different values of the
polarization, $P$, determine the temperature, $T_c$, at which the peak
in $n_{\rm pair}(k)$ shifts to $k=0$.  Figure~\ref{fig:PhaseDiag}
shows the resulting phase diagrams for $N=L$ (half filling) and
$N=L/2$ (quarter filling). We mention again that the phase boundaries
represent crossover behaviour, not phase transitions since this
one-dimensional quantum system does not have phase transitions at
finite temperature. The Fermi temperatures shown in the figure are
calculated assuming equal populations using $\epsilon_F=tk_F^2$ where
$\epsilon_F$ is the Fermi energy.

The phase diagrams show clearly that the FFLO phase is quite robust,
persisting over a wide range of polarizations and to rather high
temperatures. For $N=L$, it persists up to $T/T_F\approx 0.2$ and for
$N=L/2$ up to $T/T_F\approx 0.8$. We also see that, in both cases, the
crossover temperature increases with the polarization up to a maximum
value after which it decreases again.  This can be understood
physically as follows: When $P$ is small, the Fermi ``surfaces'' of
the two populations are so close to matching that very little thermal
energy is needed to get them to match. Thus even at very low finite
temperature, pairing takes place at zero center-of-mass
momentum. Therefore, larger polarizations have a stabilizing effect on
the FFLO phase.

There are numerical difficulties with the determination of the phase
diagram at very low and very high polarizations. At very high
polarization, there is a very small number of minority particles,
making the FFLO signal difficult to discern clearly. We were therefore
not able to examine $P>0.9$ in our simulations. For that reason, some
symbols on the phase boundary at high polarization are open, indicating
that up to this polarization, the system is in the FFLO phase. The
solid symbols demark the true boundary between FFLO and PPP.

In addition, at small polarization, very low temperature is needed
to observe the FFLO phase. However, there is a practical limit on how
low a temperature we can simulate since the lower the temperature the
longer the simulation time needed to obtain precise results. The dashed
(green) line connecting the origin to the first numerical points
simply schematizes the expected position of the boundary.

A phase diagram in the $(P,T)$ plane was calculated in
Ref. \cite{koponen} using mean field theory (MFT). The general shape of
the FFLO phase obtained there (Fig.~9 of Ref. \cite{koponen}) is
similar to what we found here. However there are very important
differences. For example, unlike MFT, we have found that there is no
direct transition from the FFLO phase to the Fermi liquid phase
(broken pairs): The FFLO phase is destroyed at a temperature which is
much lower than that required to break the pairs and is replaced by
the PPP.

\begin{figure}[h]
\begin{center}
  \includegraphics[width=0.45\textwidth,angle=0,clip]{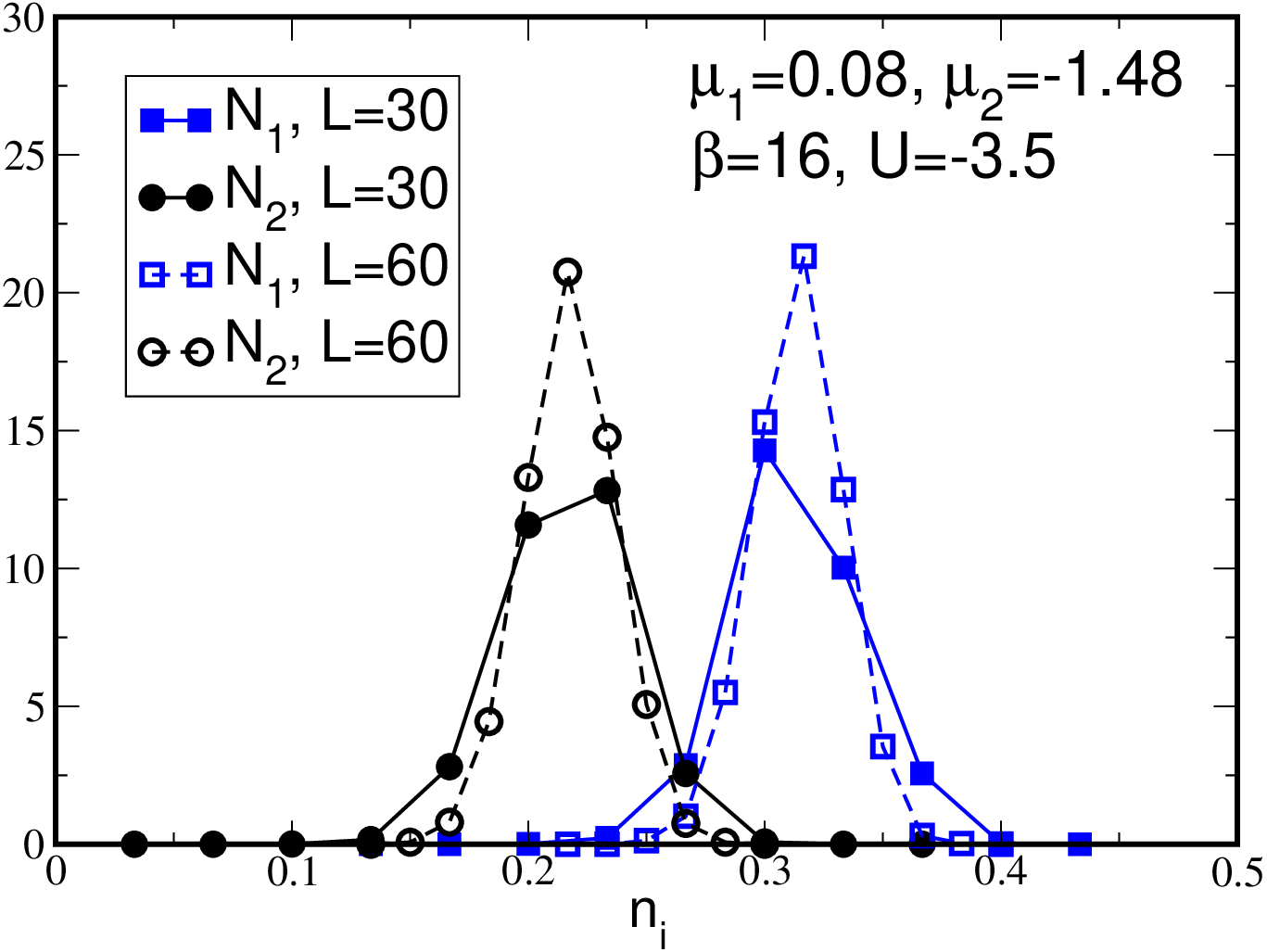}
\end{center}
\vspace{-0.6cm}
\caption{\label{fig:histogramFS} (color online). Majority ($N_1$) and
  minority ($N_2$) density histograms at the FFLO-PPP boundary for two system
sizes. The larger system size offers more grid points and, therefore, a finer
resolution of the density fluctuations. A single peak is seen for each
population indicating the absence of phase separation.
}
\end{figure}

\begin{figure}[h]
\begin{center}
\includegraphics[width=0.45\textwidth,angle=0,clip]{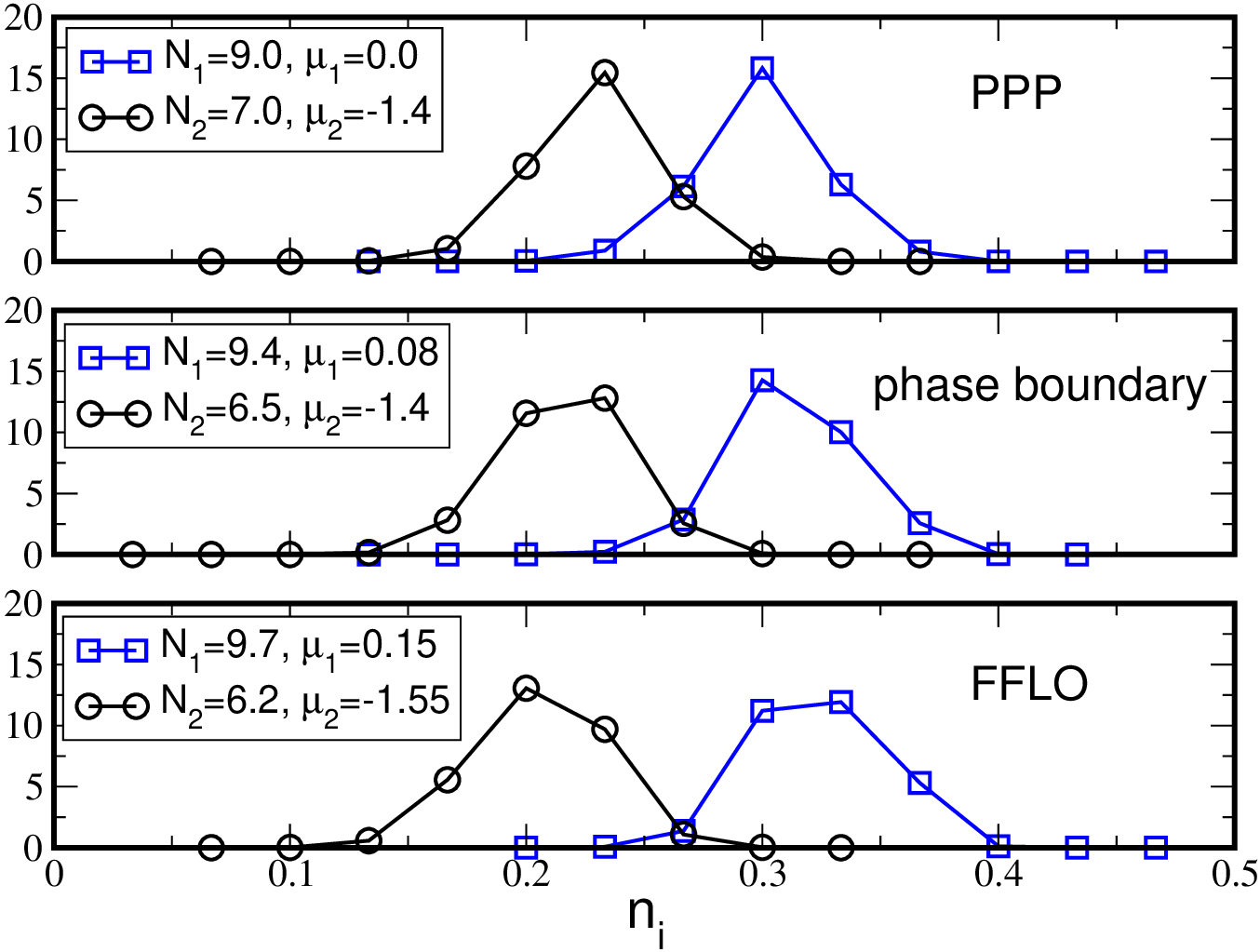}
\end{center}
\vspace{-0.6cm}
\caption{\label{fig:histogram} (color online). Histograms of local
densities for $L=30$, $\beta=16t$ and $U=-3.5t$. Left peaks correspond
to the minority species and right ones to the majority. The peaks move
smoothly as $\mu_1$ and $\mu_2$ are tuned to take the system across
the boundary between PPP and FFLO. No double peak structure is observed,
indicating the absence of phase separation.
}
\end{figure}

Reference \cite{koponen} predicts phase separation between the FFLO
and PPP at low $P$ and $T$ (Fig.~9 in Ref. \cite{koponen}). In order
to examine this possibility, we study the density histograms in the
grand canonical ensemble. The idea is as follows: Starting in, say,
the PPP, we increase the polarization by tuning the chemical
potentials, $\mu_1$ and $\mu_2$. For each choice of $\mu_1$ and
$\mu_2$, we accumulate the histograms of the particle populations. If
phase separation is present, then as the system approaches the phase
separation region, the density histogram of {\it each} species should
develop a double peak structure. If no such structure develops, it
means that there is no phase separation as the system crosses from the
PPP to the FFLO. We first verify the correct behaviour of the obtained
histogram as the size of the system is changed. In
Fig.~\ref{fig:histogramFS} we show the histograms for two system sizes
at half filling but with all other parameters fixed. We see that the
histograms for the two system sizes agree very well; the main
difference is that the larger system size (obviously) allows for a
finer grid of densities which redistributes the values a little and
exhibits the main peak more clearly. In Fig.~\ref{fig:histogram} we
show, for fixed inverse temperature $\beta=16$, the histograms for
three cases at half filling: In the top panel the system is just
inside the PPP phase, the middle panel the system is at the PPP-FFLO
boundary and the bottom panel the system is just inside the FFLO
phase. No double peak structure develops, which leads us to conclude
that there is no phase separation. This was done for several
temperatures at low polarization.

It is useful here to comment on the algorithm choice for calculating
the histograms. Although the DQMC algorithm is grand canonical and
thus allows for particle number fluctuations, it is not useful for
calculating the density histograms. The reason is that in DQMC one
changes the realization of the auxiliary Hubbard-Stratonovich field;
but for each such realization, the fermions have been traced over all
their possible configurations. On the contrary, in the grand canonical
version of the SGF algorithm, the update is done over the fermion
configurations themselves. So, the particle number can be measured
configuration by configuration.


\section{Trapped system at finite temperature}

Continuing earlier work in higher dimension
\cite{zwierlein06,partridge06}, the Rice group \cite{hulet} recently
reported on experiments in one dimensional confined Fermi systems
($^6Li$ atoms) with imbalanced populations. These experiments were
done in the continuum, {\it i.e.}~without an optical lattice, and
focused on the behaviour of the system in three polarization regimes
by measuring the density profiles of the fermionic species. It was
found that the central part of the system is always partially
polarized whereas the behaviour of the outlying regions depends on the
total polarization. For low polarization, $P=0.05$, the outlying
regions were found to be fully paired in the sense that the density
profiles of the two fermion species matched within experimental
precision. For medium polarization, $P=0.15$, the density profiles
indicated that the whole system is partially polarized. Finally, for
large polarization, $P=0.59$, the wings were found to be populated
exclusively by the excess fermion species and thus were fully
polarized.

The experiment \cite{hulet} consisted of a two-dimensional array of
elongated (one-dimensional) tubes. Along the tube, the axial
direction, the atoms were confined with a trap frequency
$\omega_z=2\pi\times 200$Hz; in the central tube, the total number of
atoms at zero polarization was approximately $250$ and the temperature
was estimated at $T/T_F\approx 0.1$. The pair binding energy,
$\epsilon=\hbar^2/ma_{1D}$ (where $a_{1D}$ is the effective
one-dimensional scattering length), was estimated to be
$\epsilon/\epsilon_F\approx 5.3 $ with the Fermi energy calculated
assuming a balanced system with a total of $250$ particles.

In this section, we present QMC results for the fermionic Hubbard
model, Eq.(\ref{Hamiltonian}), in the presence of the confining
trap. Our goal is to make contact with the above mentioned experiment
in the continuum; to this end we simulate lattice systems that are
dilute enough so that the fermions in the center of the trap are far
from forming a flat plateau corresponding to a band insulator.  We
introduce the trapping potential in Eq.(\ref{Hamiltonian})
$V_T=0.0007t$ which corresponds to $\hbar\omega_z=2\sqrt{tV_T}$. The
total number of particles in our simulations for balanced populations
is $78$, to be compared with $250$ in the experiment. We performed our
simulations in the temperature range $0.008\leq T/T_F\leq 0.25$ which
includes the temperature at which the experiments were performed,
$T/T_F=0.1$. In addition, to place our system in the same coupling
parameter regime as the experiments, we present our results for a
coupling strength of $U=-10t$. $U$ is the ``pair binding energy'' and
the value we have chosen gives $|U|/\epsilon_F=4.8$, close to the
experimental value.

\begin{figure}[h]
\begin{center}
  \includegraphics[width=0.45\textwidth,angle=0,clip]{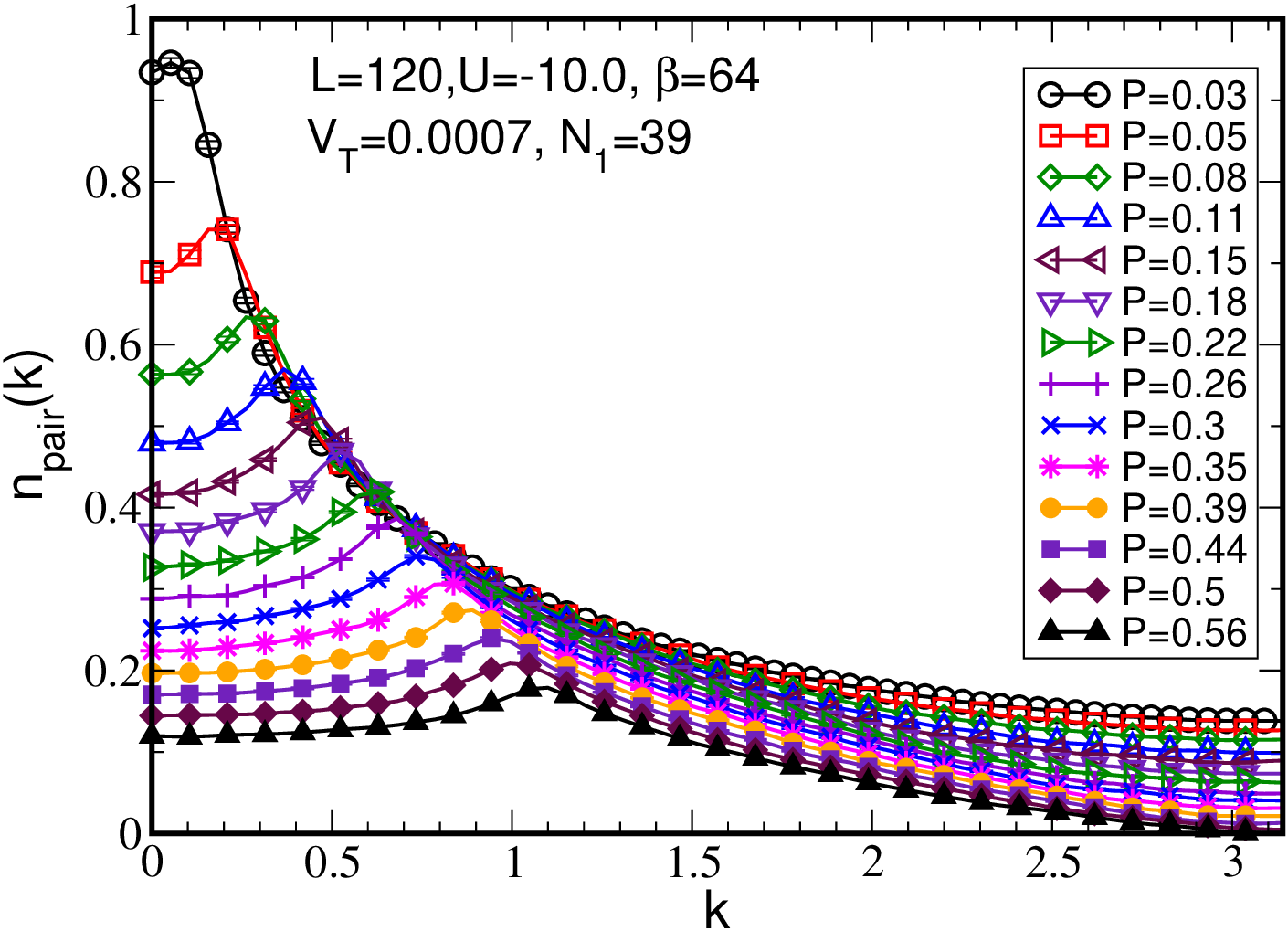}
\end{center}
\vspace{-0.6cm}
\caption{\label{fig:trapScanP} (color online). Pair momentum
distribution in a trapped system for several polarizations.  The FFLO
peak moves to higher momentum values as $P$ increases as in the
uniform system.  The majority population is fixed $N_1=39$,
$T/T_F=0.008$ in the balanced case $N_1=N_2=39$.} 
\end{figure}

First we look at the system at low temperature when we increase the
polarization. As in the homogenous case, the pair momentum
distribution exhibits a maximum at $k_{\rm peak}\neq0$
(Fig.~\ref{fig:trapScanP}) and as before $k_{\rm peak}$ increases with
growing polarization. Looking at the density profiles one immediately
observes that the low and high polarization regimes differ
significantly.

\subsection{Low polarization}
\begin{figure}[h]
\begin{center}
  \includegraphics[width=0.45\textwidth,angle=0,clip]{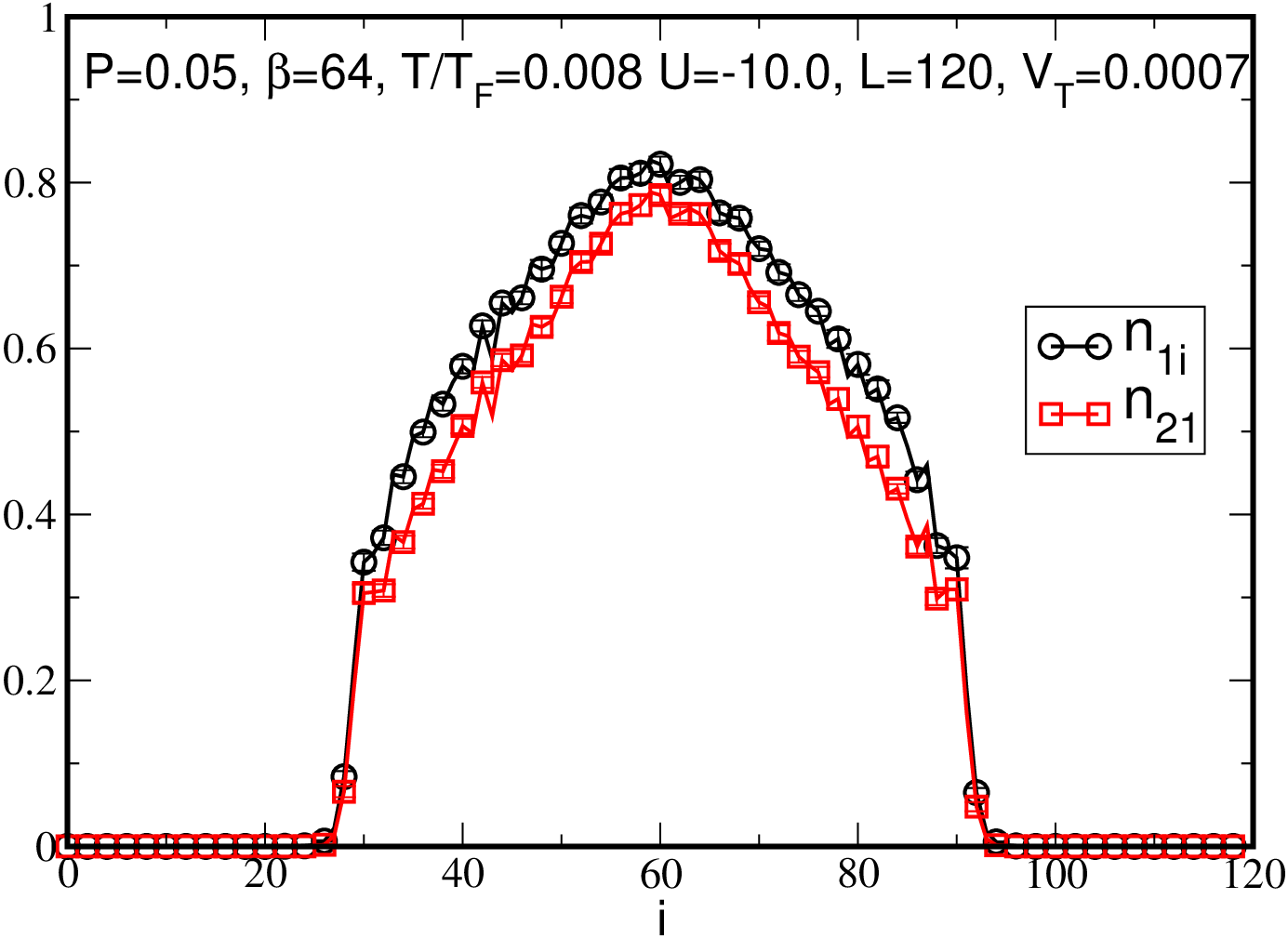}
\end{center}
\vspace{-0.6cm}
\caption{\label{fig:densityprofilelowPlowT} (color online). Density
  profiles of the two species for the low temperature and low
  polarization case. Very narrow fully paired regions are seen at the
  edges of the cloud. $N_1=39$ and $N_2=35$.}
\end{figure}

In Fig.~\ref{fig:densityprofilelowPlowT} we show the density profiles
at very low temperature, $T/T_F=0.008$, for a system at very small
polarization, $P=0.05$, corresponding to the open (red) squares in
Fig.~\ref{fig:trapScanP}. The central region of the system is clearly
partially polarized: the profiles do not overlap. This partial
polarization, {\it i.e.} population imbalance, causes FFLO pairing to
take place as is evidenced by the pair momentum distribution in
Fig.~\ref{fig:trapScanP}.  However, in a very narrow interval at the
edges of the system, the density profiles match very closely and the
system is fully paired \cite{DrummondfiniteT,hulet}. This fully paired
region was studied in the continuum \cite{casula} using Path Integral
Monte Carlo (PIMC) simulations. It was shown in Ref. \cite{casula} that
at a polarization $P=0.04$, the fully paired region exists for
$T<0.025 T_F$ and $T<0.035 T_F$ for the two strong couplings studied
(the first value corresponds to the smaller coupling). The narrowness
of this region on the lattice was studied at $T=0$ in
Ref. \cite{Meisner,tezuka10}.

\begin{figure}[h]
\begin{center}
  \includegraphics[width=0.45\textwidth,angle=0,clip]{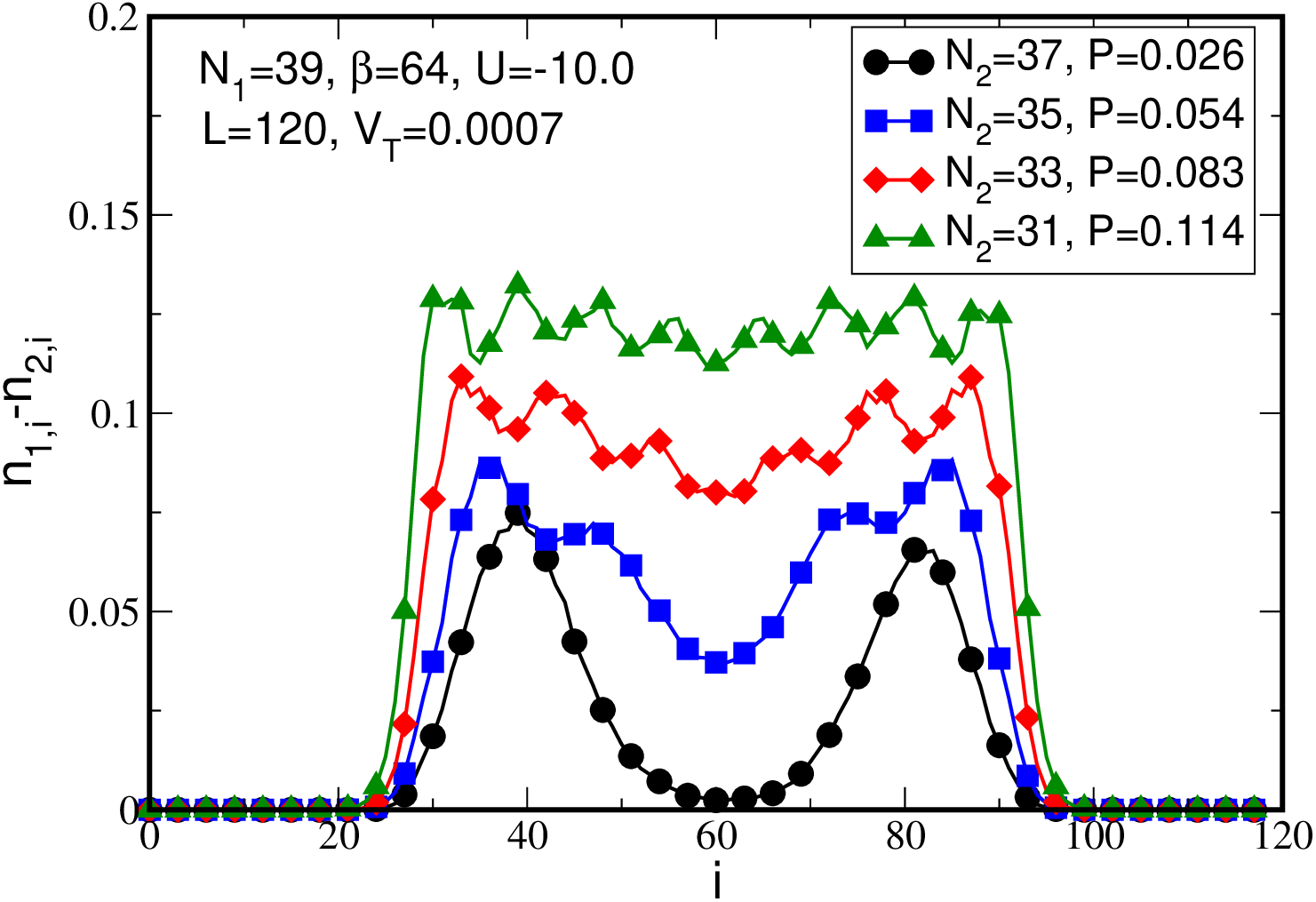}
\end{center}
\vspace{-0.6cm}
\caption{\label{fig:differencesLowT} (color online). Density profile
  differences for different polarizations. The oscillations are
  standing waves showing the pair rich (minima) and pair depleted
  (maxima) regions in the confined system.  $T/T_F=0.008$ for the
  balanced case with $N=78$ particles.}
\end{figure}

The difference in the density profiles,
Fig.~\ref{fig:differencesLowT}, of the two species shows regular
oscillations indicating that the partially polarized region is not
uniform. Such oscillations have been seen before
\cite{feiguin07,tezuka07,batrouni08}. They correspond to the standing
wave of length $\lambda =2\pi/k_{\rm peak}$, as can be easily verified
from Figs.~\ref{fig:trapScanP} and \ref{fig:differencesLowT}, and
describe the length scale at which the system passes from pair-rich to
pair-poor regions. This is a striking visual demonstration that the
FFLO phase is not uniform.

\begin{figure}[h]
\begin{center}
  \includegraphics[width=0.45\textwidth,angle=0,clip]{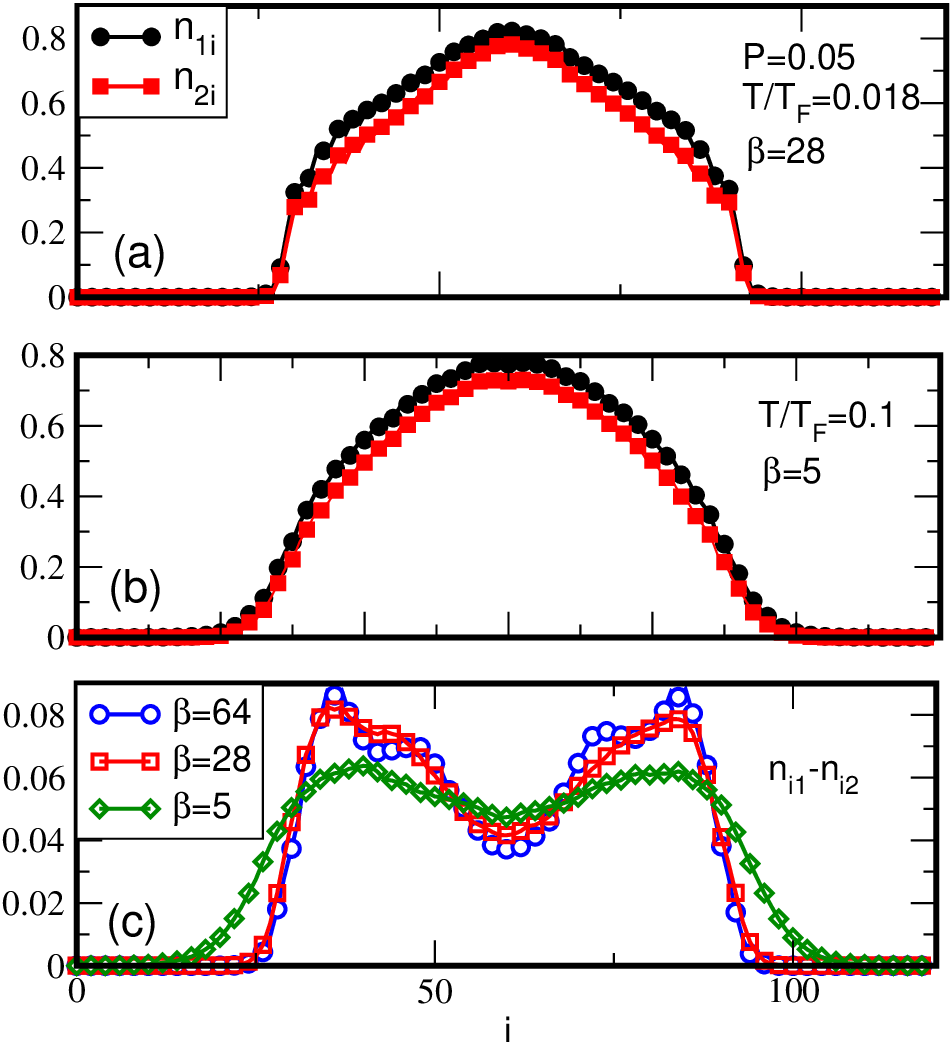}
\end{center}

\vspace{-0.6cm}
\caption{\label{fig:traplowPScanB} (color online). Trapped system at
  finite temperature and low polarization.  (a) and (b) density
  profiles at $T/T_F=0.016$ and $T/T_F=0.1$ respectively. (c) density
  profile difference for different temperatures.}
\end{figure}

\begin{figure}[h]
\begin{center}
\includegraphics[width=0.40\textwidth,angle=0,clip]{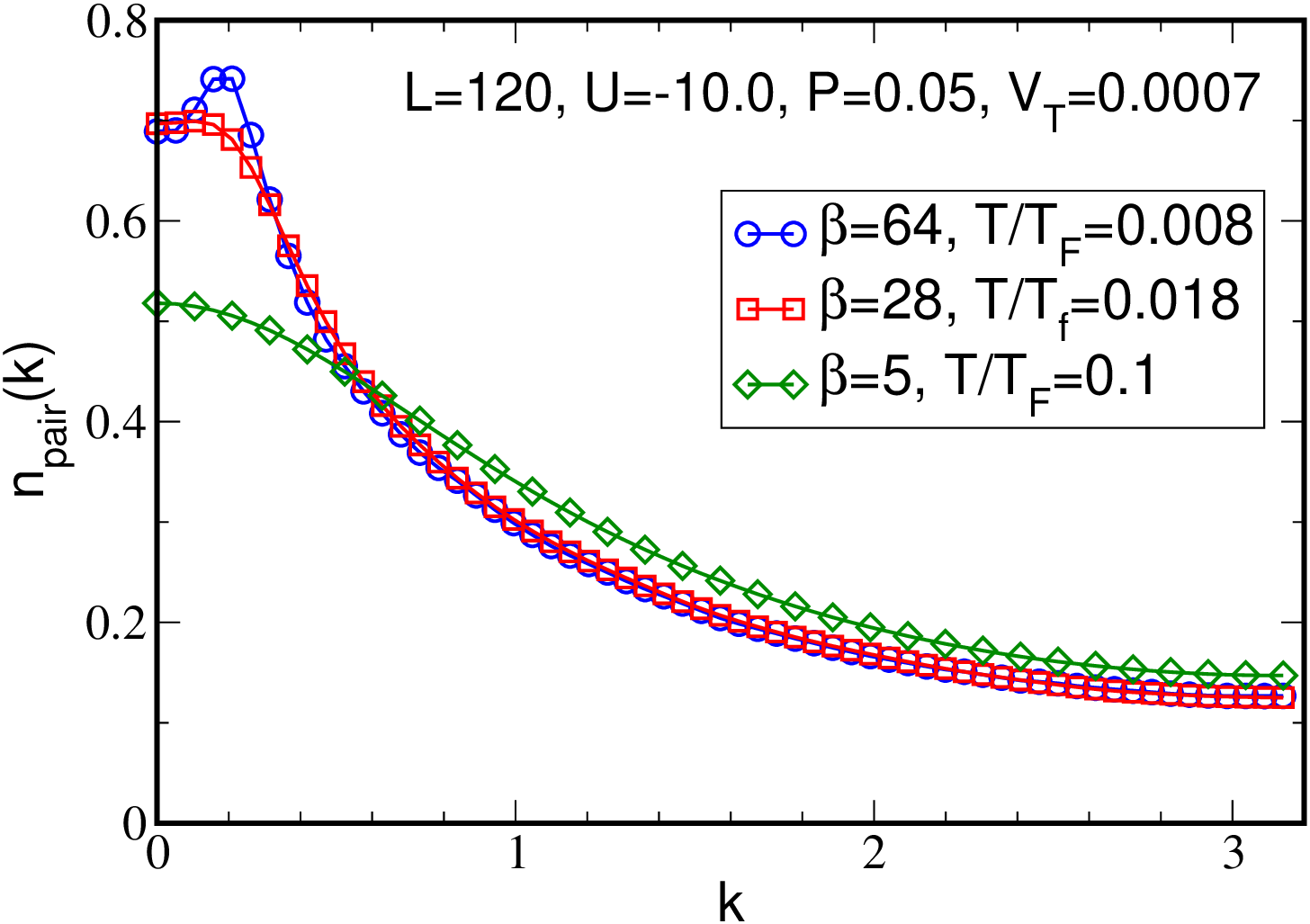}
\end{center}
\vspace{-0.6cm}
\caption{\label{fig:traplowPScanBpairmom} (color online). Pair
momentum distribution of the system at P=0.05 (Fig. \ref{fig:traplowPScanB})
and increasing temperature. The FFLO peak disappears at $T/T_F=0.016$.}
\end{figure}

Next, we examine temperature effects on the system at low
polarization. The system at $\beta=64$,
Fig.~\ref{fig:densityprofilelowPlowT}, is now heated to $\beta =28$
($T/T_F=0.016$) and $\beta=5$ ($T/T_F=0.1$) as shown in
Fig.~\ref{fig:traplowPScanB} (a) and (b) respectively.  As the
temperature increases, the clouds spread out and the profiles become
more rounded. Figure \ref{fig:traplowPScanB} (c) shows what happens to
the population differences as the temperature rises. As $T$ increases,
the population difference vanishes in the wings more gradually than at
the lower temperatures. In addition, the oscillations which indicate
the presence of FFLO get smoothed out substantially at $\beta=28$ and
have essentially disappeared for $\beta=5$. This is confirmed by the
behaviour of the pair momentum distribution, displayed in
Fig.~\ref{fig:traplowPScanBpairmom}, which shows the FFLO peak
disappearing as $T$ is increased to $T=0.016T_F$.  This value is
smaller than, but consistent with, the phase diagram in
Ref. \cite{DrummondfiniteT} (Fig. 1) which shows that at $P=0.05$ the
FFLO phase disappears for $T > 0.05T_F$. Our value is also consistent
with that found in Ref. \cite{casula}.  This illustrates the fragility
of the fully paired paired wings and the FFLO phase at low
polarization.

\subsection{High polarization}\

In section III, we calculated the phase diagram of the uniform system
and showed that the FFLO phase persists to higher temperatures for
larger polarization. We now demonstrate the same effect in the trapped
system.

\begin{figure}[h]
\begin{center}
\includegraphics[width=0.40\textwidth,angle=0,clip]{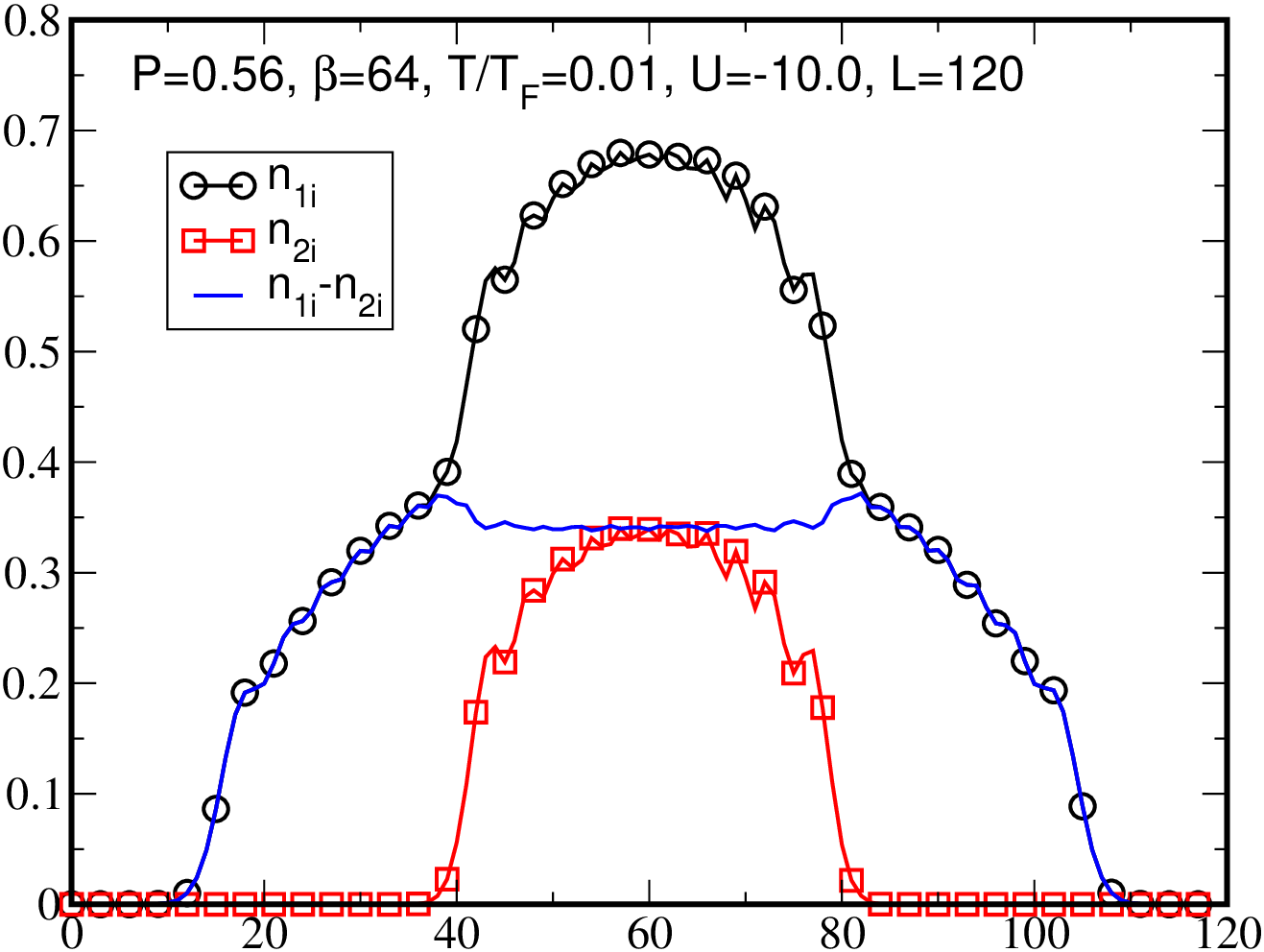}
\end{center}

\vspace{-0.6cm}
\caption{\label{fig:traphighPlowB} (color online). Trapped system at
  low temperature and high polarization. $N_1=39, N_2=11, L=120$,
  $U/t=-10$, $V_t=0.0007$. $T_F$ is calculated for the total balanced
  population of $N=50$. Large fully polarized regions are seen at the
boundaries of the system.}
\end{figure}

\begin{figure}[h]
\begin{center}
\includegraphics[width=0.45\textwidth,angle=0,clip]{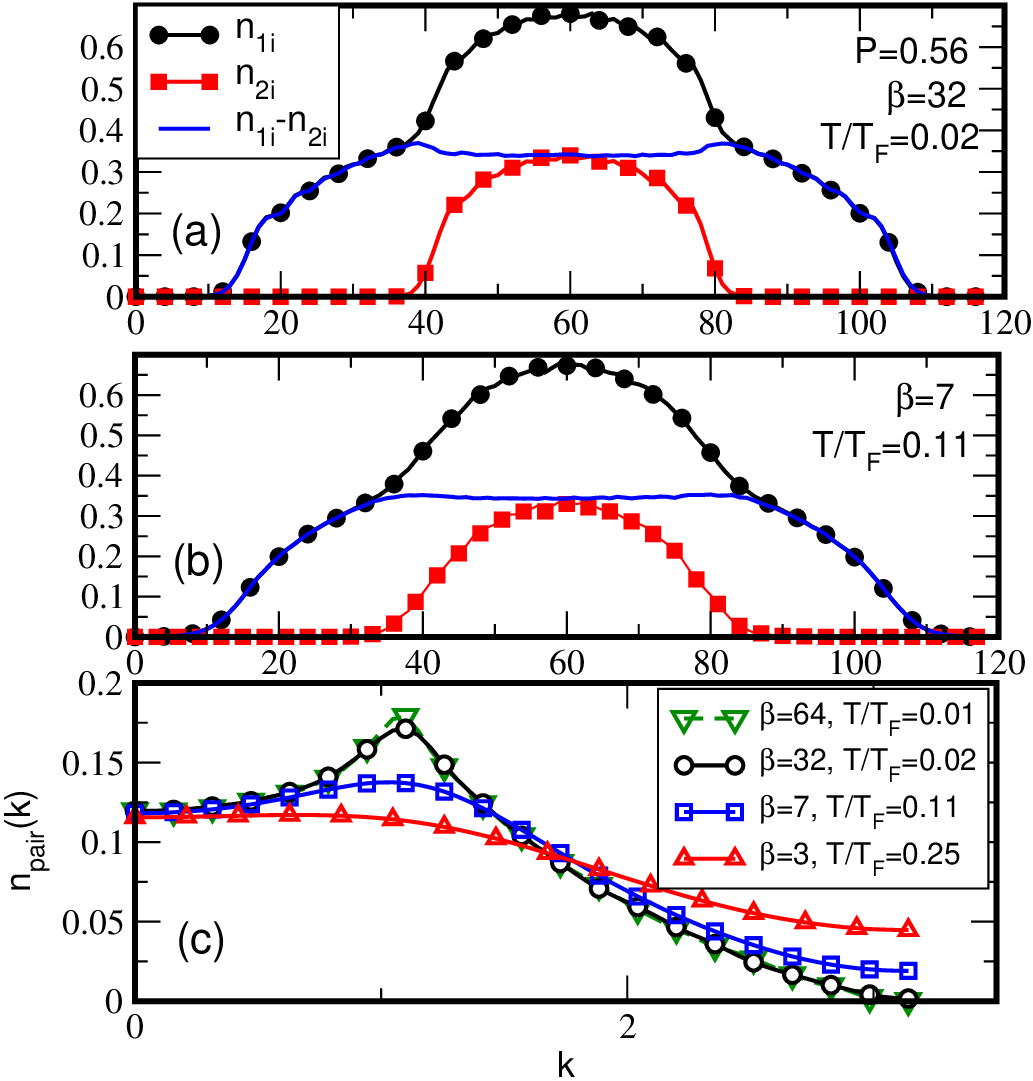}
\end{center}

\vspace{-0.6cm}
\caption{\label{fig:traphighPScanB} (color online). Same trapped
  system as in Fig.~\ref{fig:traphighPlowB} at higher temperature.
  $P=0.56, L=120$, $U/t=-10$, $V_t=0.0007t$, $N_1=39$, $N_2=11$. (a)
  and (b) show the density profiles and differences for two
  temperature. (c) Pair momentum distribution.}
\end{figure}

Figure \ref{fig:traphighPlowB} shows the density profiles in the
system with $P=0.56$ at $\beta=64$ ($T/T_F=0.01$). It is clear at this
large $P$ that the central region is partially polarized and the wings
are fully polarized, populated only by the majority species as was
also observed in \cite{feiguin07}. The population difference in the
central region of the system in Fig.~\ref{fig:traphighPlowB} is almost
constant but with oscillations, $\lambda=2\pi/k_{\rm peak}$, due to
the presence of FFLO pairing as shown clearly by the pair momentum
distribution in Fig.~\ref{fig:traphighPScanB} (c).

As the temperature is increased, this shell structure, {\it
  i.e.}~partially polarized core and fully polarized wings, persists,
as can be seen in Fig.~\ref{fig:traphighPScanB} (a) and (b). However,
we observe that the partially polarized core expands in size and the
fully polarized population in the wings decreases. We also see that
the FFLO phase at this large polarization is stabilized significantly
compared with the $P=0.05$ case
(Fig.~\ref{fig:traplowPScanBpairmom}). For $P=0.56$, the FFLO peak
persists, albeit weakly, up to $T/T_F=0.11$, vanishing completely at
$T/T_F=0.25$. This result is encouraging for the experiments
\cite{hulet} which can be done at high polarization and $T/T_F\approx
0.1$. However, at this temperature, the FFLO peak is not very
pronounced and might still be difficult to observe experimentally.

We note that, as in the case of the uniform system, the FFLO phase
disappears at a temperature which is much lower than the contact
potential energy. In the case above, the FFLO phase disappears at
$\beta t\approx 3$ while the contact interactions is $U=-10t$. The
approximate condition to break the pairs is $\beta U\approx 1$; we see
that $\beta t=3$ is not a high enough temperature to break the pairs.
Therefore, as $T$ is increased, the FFLO phase is replaced by the PPP
discussed in the previous section and not by the normal state composed
of the two unpaired spin populations. Our QMC result is in
disagreement with mean field predictions that as $T$ is increased for
high polarization, the system passes directly from the FFLO phase to
the normal state \cite{DrummondfiniteT}.

As in the uniform case, the question arises as to whether one can
stabilize the FFLO phase at higher temperature simply by increasing
the attractive interactcion. The answer is the same as in the uniform
case: As the attractive interaction is increased, the FFLO peak first
increases in height but then saturates and starts to decrease.  In
addition, as $|U|$ is increased, the partially polarized core region
shrinks and the polarization in that zone increases. This has the
effect of shifting the FFLO peak to larger values of $k$.
For the fillings we conisdered here, the value of the interactions
we took, $U=-10t$, appears to be near optimal.


\section{Conclusions}

In this paper, we used three QMC algorithms (DQMC, canonical SGF and
grand canonical SGF) to study the behaviour of one-dimensional
imbalanced fermion systems with attractive interactions governed by
the Hubbard Hamiltonian.  We explored both the uniform and the
confined cases.

In the uniform situation, we mapped the phase diagram in the
polarization-temperature plane at two values of the total density
(half filling and quarter filling) at the same interaction strength
$U=-3.5t$. Both cases show the same general features: (1) The FFLO
phase is very robust in the ground state and exists for a very wide
range of polarizations; (2) at small polarization, the FFLO phase is
very sensitive to temperature increase; (3) the FFLO phase persists to
higher temperature when the polarization is larger and (4) the
temperature at which the FFLO phase disappears is not high enough to
break up the pairs leading to a spatially uniform polarized paired
phase.

In the confined case, we performed our simulations for system
parameters comparable to those in the experiment of Ref. \cite{hulet}.
We found the stability features of FFLO in the confined phase to be
similar to those in the uniform case.  At low polarization, the FFLO
phase is destroyed even for a temperature as low as
$T/T_F=0.016$. However, and this is significant for experimental
efforts to detect FFLO in the confined system, we found that at large
polarization, the FFLO phase persists at $T/T_F>0.11$ a temperature
higher than that of the experiment, which has $T/T_F\approx
0.1$. Finally, the temperature at which the FFLO phase is destroyed is
not high enough to break the pairs in disagreement with mean field
results \cite{DrummondfiniteT}.

\begin{acknowledgments}
This work was supported by: an ARO Award W911NF0710576 with funds from
the DARPA OLE Program; by the CNRS-UC Davis EPOCAL joint research
grant; by NSF grant OISE-0952300; by the France-Singapore Merlion
program (PHC Egide, SpinCold 2.02.07 and FermiCold 2.01.09) and the
CNRS PICS 4159 (France). Centre for Quantum Technologies is a Research
Centre of Excellence funded by the Ministry of Education and the
National Research Foundation of Singapore.  We thank R. K. Kid for
helpful insight.

\end{acknowledgments} 

{}

\end{document}